\documentclass[usenatbib]{mn2e}
\usepackage{psfig}
\title{The morphology-density relation in X-ray bright galaxy groups}
\author[S. F. Helsdon \& T. J. Ponman]
        {Stephen F. Helsdon$^{1,2}$\thanks{E-mail: sfh@ociw.edu} and Trevor
         J. Ponman$^{2}$\\
$^1$  The Observatories of the Carnegie Institute of Washington, 813 Santa Barbara Street, Pasadena, CA 91101\\  
$^{2}$  School of Physics and Astronomy, University of
        Birmingham, Edgbaston, Birmingham B15 2TT, UK\\}
 \date{Accepted 2002 ??.
      Received 2001 ??;
      in original form 2001 ??}

\pagerange{\pageref{firstpage}--\pageref{lastpage}}
\pubyear{2001}

\begin{document}

\maketitle

\label{firstpage}

\begin{abstract}
  
  We have examined the morphological make-up of X-ray bright
  groups. The brighter galaxies in these groups exhibit clear
  morphology-density and morphology-radius relations. The group
  morphology-density relation is offset from the cluster relation in
  the sense that at a given surface density, X-ray bright groups have
  a lower spiral fraction. After correcting for projection effects the
  morphology-3D density relation is still shifted towards fewer
  spirals for a given 3D density, in comparison with clusters. A
  simple model which corrects the group data for the effects of
  projection and for the expected higher merging rate in groups,
  brings the morphology-density relation into good agreement with that
  of clusters, suggesting that the relation may be driven by two-body
  interactions.  The fraction of S0 galaxies in these X-ray bright
  groups is at least as high as that observed in nearby
  clusters. Given the low velocity dispersion of groups, this
  indicates that ram pressure stripping is not the dominant mechanism
  for S0 formation.
  
\end{abstract}

%%%%%%%%%%%%%%%%%%%%%%%%%%%%%%%%%%%%%%%


%%%%%%%%%%%%%%%%%%%%%%%%%%%%%%%%%%%%%%%

\section{Introduction}
\label{sec:intro}

The well known relationship between galaxy morphology and density (e.g.
\citealt{dressler80,dressler97}) suggests that the environment of a galaxy
may have a significant effect on its evolution. By examining the
properties of galaxies as a function of environment it should be
possible to gain important insights into the evolution of both
galaxies and galaxy systems.  Particularly interesting is the group
environment, which is typically made up of between three and a few
tens of galaxies. These systems are gravitationally bound systems in
which the density and velocities of the member galaxies suggest that
mergers and interactions are more common than in clusters or in the
field (e.g. \citealt{mamon92,mamon00}). In addition, the majority of
galaxies are found in a group environment \citep{tully87}, and many
galaxies in clusters will once have been part of groups.

Perhaps the most striking connection between group and galaxy properties is
the fact that almost all X-ray bright groups contain a bright
early-type galaxy located at the spatial and kinematic centre of the group
\citep{zabludoff98}. This central galaxy is presumably the result of early
merging activity in the collapsing group core \citep{governato96} and
given the high expected rates of mergers in present day groups, this
central galaxy may still be rapidly growing through mergers and
accretion. Other galaxies in the group are also likely to be affected
by the group environment. \citet{postman84} and \citet{ramella99} have
shown that galaxies in groups follow a morphology-density
relation. However, it is not clear if this is the same relation as for
clusters as some X-ray bright groups appear to have spiral fractions
consistent with those found in rich clusters \citep{zabludoff98}.

\begin{figure*}
\hspace{0cm}
\psfig{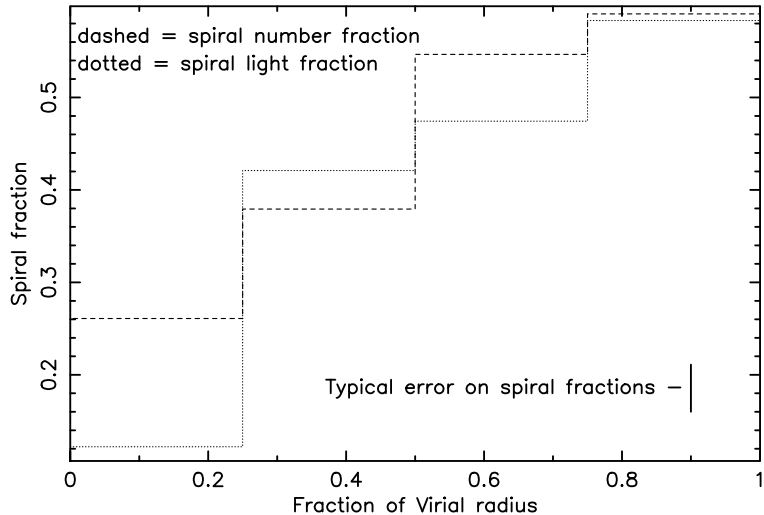}
\caption{\label{fig:spiral_radius}Spiral fraction (including irregulars) 
versus radius (in units of the virial radius, $R_V$) for X-ray bright
groups. Each radial bin contains at least 75 galaxies.}
\end{figure*} 

It is also known that there is a connection between the morphological
make-up of groups and the presence of diffuse X-ray emission (e.g.
\citealt{pildis95,henry95,mulchaey96}), in the sense that X-ray detected
groups tend to be spiral poor, whilst undetected groups contain more
late-types. In this and a companion paper \citep{helsdon02} we aim to
look in some detail at the relationship between the X-ray and optical
properties of X-ray bright galaxy groups. Here we focus on the
morphological makeup of these systems, and examine the
morphology-density relation of X-ray bright groups.

%%%%%%%%%%%%%%%%%%%%%%%%%%%%%%%%%%%%%%%

\section{The Sample}
\label{sec:sample}

The group sample used here is the sample of 24 X-ray bright groups
originally studied by \citep{helsdon00}. The X-ray properties of this
sample are discussed extensively in \citet{helsdon00} and
\citet{helsdon00b}, but the important thing for the purposes of this 
work is that all these groups have detected, extended diffuse
emission, associated with the group. The presence of a hot intragroup
medium indicates that the group is a real gravitationally bound
object, rather than just a chance superposition of a few galaxies ---
potentially a serious problem with optically selected groups.

We also need to determine the galaxy memberships of these
groups. Unfortunately, it is not satisfactory to simply use the galaxy
memberships as given in the original group catalogue, since the
galaxies were originally selected from different sources which in
general have different selection criteria. This problem is addressed
in \citet{helsdon02}, and here we use the same technique. This
procedure is summarised below --- for full details see
\citet{helsdon02}. For each group we search the NASA/IPAC Extragalactic 
Database (NED) and the Lyon-Meudon Extragalactic Database (LEDA) for
galaxies lying within the group virial radius in projection on the
sky, and having recession velocities within three times the group
velocity dispersion from the catalogued group mean. The centre
position used for each group is the centre of the X-ray emission. In
most cases this position is very close to the position of the central
galaxy, with the three exceptions being bimodal systems in which the
X-ray centre falls roughly between the two main galaxies. It is
possible that the centres used in the bimodal groups could act to
obscure radial trends present in the other groups. However, as
we average over all groups when looking at radial trends, and the
bimodal systems only make up a small fraction of the sample, this is
unlikely to have a significant effect on any results.

To ensure an even luminosity cut in all groups we use the galaxy
luminosity function of \citet{zabludoff00} (for X-ray bright groups)
to employ two magnitude cuts --- one which should include 50\% of the
total optical group light and one which should include 90\% of the
total group light. For the 50\% cut \citet{helsdon02} show that for
the typical group in this sample the membership should be almost 100\%
complete, whilst approximately 35\% of the total group light is
missing from the 90\% cut sample. Morphological types are taken
firstly from NED if available, and then LEDA.

It should be noted that the group sample used here should not be
regarded as being statistically complete in any way. However, we do
not believe that this will introduce any particular bias, other than
the fact that since we only use groups with detected diffuse X-ray
emission, we do not include systems with undetectably faint
intergalactic gas. The group sample should rather be regarded as a
reasonably representative sample of X-ray bright groups.

%%%%%%%%%%%%%%%%%%%%%%%%%%%%%%%%%%%%%%%
\begin{figure*}
\hspace{0cm}
\psfig{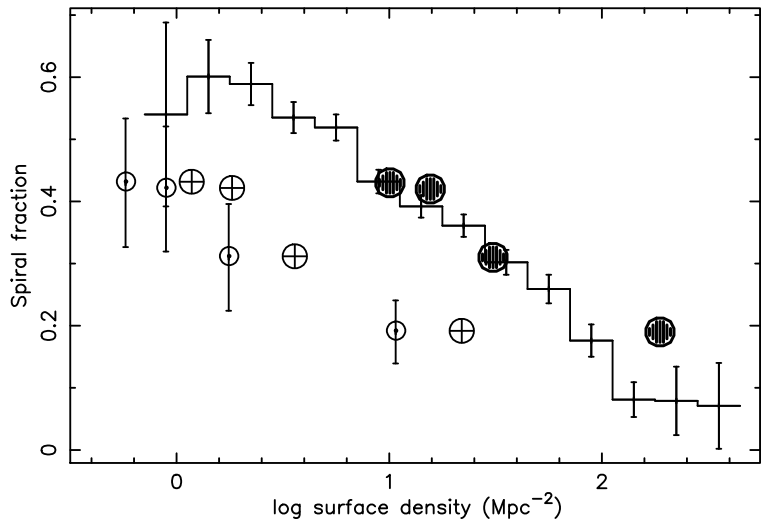}
\caption{\label{fig:spiral_rho}Spiral fraction (including irregulars) 
as a function of surface density (averaged over radial
bins). Histogram is the relation for local clusters
\protect\citep{dressler97}, and the open circles with error bars are 
the group data. Crossed and shaded circles represent group points
after estimated corrections for 3d density, and merging rates
respectively (for clarity the error bars are omitted from these
points).}
\end{figure*}

\section{The morphology-density relation}

The overall spiral fraction of all these X-ray bright groups combined
is $0.425 \pm 0.035$, however this spiral fraction drops to $0.30 \pm
0.05$ if the sample is restricted to galaxies brighter than the 50\%
luminosity cut.  Thus these groups have lower spiral fractions than
found in the field ($\sim$ 0.6 --- \citealt{postman84,whitmore93}),
particularly amongst the more luminous galaxies.
 
The spiral fraction also appears to increase with radius in these
systems.  In Figure~\ref{fig:spiral_radius} we plot the spiral
fraction calculated from all galaxies in all groups, as a function of
radius (in units of the virial radius). This plot shows that the
spiral fraction drops from about 0.6 at the outer edge (consistent
with the field population) to a value of about 0.2 in the innermost
regions of the groups. The spiral number fraction and the spiral light
fraction are similar to one another until the central bin, where the
spiral light fraction drops to less than half the number fraction, due
to the presence of a central bright early type in almost all these
groups.  The spiral fraction profile for the 50\% sample is similar to
the 90\% sample shown, but has a slightly lower spiral fraction at
each point.

This morphology-radius trend is similar to that observed in galaxy
clusters \citep{whitmore93} and also reported in X-ray bright groups
\citep{tran01}. We have also looked for evidence of a
morphology-density relation. We cannot calculate a local surface
density for each galaxy in our sample as has been done in previous
cluster work (e.g. \citealt{dressler80,dressler97}) as we do not have
sufficient numbers of galaxies in each group. Instead we calculate the
average density of each of the 4 radial bins plotted on
Figure~\ref{fig:spiral_radius}. We plot these points in
Figure~\ref{fig:spiral_rho} along with data from the local cluster
sample of \citet{dressler97}. Note that the 50\% group sample data is
used, as the magnitude cut for this sample is closest to the average
of that used by \citeauthor{dressler97} for the local cluster
data (the local cluster data is derived from an apparent magnitude cut
of $V=16.5$, which in turn corresponds to an absolute magnitude of
$M_V=-20.5$ at the average cluster distance. Our 50\% group sample
B-band cut falls within the range of magnitude cuts for the local
cluster sample and is between 0.5 and 0.9 magnitudes brighter than the
average, depending on galaxy type).

\begin{figure*}
\hspace{0cm}
\psfig{file=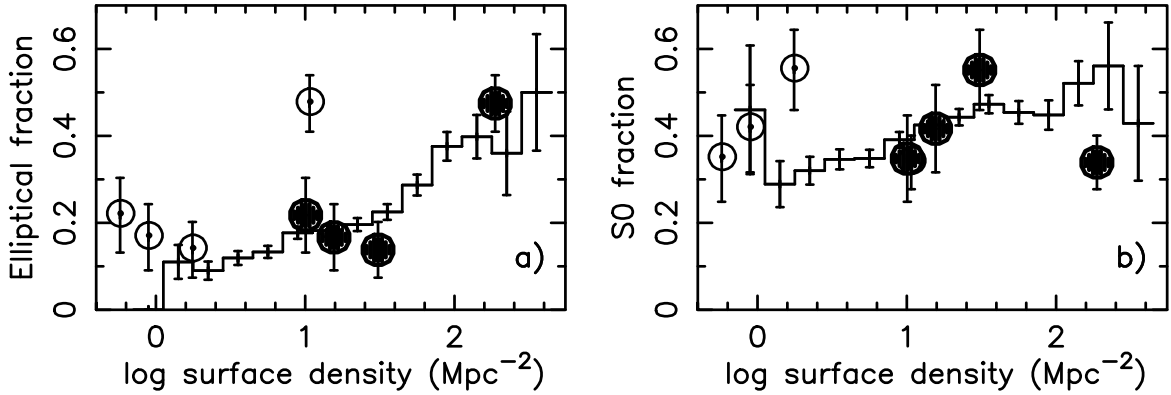,angle=-90,width=16cm}
\caption{\label{fig:e_s0_rho}a) Elliptical and b) S0 fractions as a 
  function of surface density. The histograms are the relations for
  local clusters from \protect\citet{dressler97}, open circles are the
  group data and shaded circles are the group points after corrections
  to the surface density for the difference in merging rate between
  the group and cluster data.}
\end{figure*}

As can be seen in Figure~\ref{fig:spiral_rho}, the group points
(circles) do indeed appear to follow a morphology-density
relation. However, it is clear that the group relation is offset from
the cluster data of \citeauthor{dressler97}.  One explanation for this
offset could be that the relation is driven by 3D density, rather than
by {\it projected} density.  In a group, fewer galaxies will be
projected along the line of sight onto the centre of the group. Thus
if a group and cluster had the same 3D density at the centre, the
projected density of the group would be lower than that of the
cluster. The projected surface density ($\Sigma$) should be related to
the 3D density ($\rho$) by, $\Sigma
\propto \rho R$, where $R$ is the radius of the cluster. Assuming
self-similarity we have, $R \propto T^{\frac{1}{2}}$, and therefore
$\Sigma \propto \rho T^{\frac{1}{2}}$. So by using a mean group and
mean cluster temperature, it is possible to correct for the effects of
projection on the group data, and compare with the cluster data.  Of
the 55 clusters in the \citeauthor{dressler97} local cluster sample
we obtain temperatures for 31 of them from \citet{white97} and use the
mean of these as the typical cluster temperature. This gives a mean
cluster temperature of $T=3.85$ keV in comparison with a mean
temperature of $T=0.92$ keV for the group sample.

The group data corrected for the effects of 3D projection are also
plotted (crossed circles) in Figure~\ref{fig:spiral_rho}. It can be
seen that the points have moved in such a way as to reduce the
discrepancy between the group and cluster points, but there is still a
significant offset between the datasets. At comparable local 3D
densities, X-ray bright groups appear to have fewer spirals and more
early-types than clusters. This result is inconsistent with that of
\citet{postman84}, who found a universal morphology density relation
for groups and clusters, with low density groups being more spiral
rich than high density clusters, and with the higher density regions
in groups having the same morphological mix as cluster regions of
comparable 3D density. However, the groups used by \citet{postman84}
were optically selected, and it is possible that they included some
spurious groups or groups at an early stage of virialisation.  Both of
these effects would increase the spiral fraction, and would therefore
move the group data towards the cluster data. In comparison, our
sample consists of collapsed, X-ray bright groups. It should also be
pointed out that the trend of fewer spirals for a given 3D density is
opposite to that noted by \citet{mamon86} and \citet{hickson88} for
the Hickson Compact Groups (HCG) \citep{hickson82}. However, the
apparent densities of HCGs are subject to complicated and subtle
selection effects (e.g. \citealt{prandoni94,mamon00b}) and in light of
this it may not be particularly surprising that they behave in a
different way.

So whilst it seems that galaxies in X-ray bright groups do follow a
morphology-density relation, this relation appears to be different to
that observed in clusters. However, groups are just small clusters,
and it is reasonable to assume that similar processes govern the
morphological makeup in both groups and clusters. The group sample
here is probing the morphology-density relation in systems
approximately an order of magnitude less massive than the clusters in
\citet{dressler97}, and the variations seen over this large mass range
suggest that an additional parameter is needed in the
morphology-density relation. 

A likely candidate for this extra parameter is the merging rate --
galaxies in groups have a higher merging rate than galaxies in
clusters, due to their lower velocity dispersions. Using different
merger cross-sections, \citet{mamon92} and \citet{makino97} have shown
that the merging rate ($k$) scales as $ k \propto
\sigma_{\rm cl}^{-3}$, where $\sigma_{\rm cl}$ is the cluster velocity
dispersion, and \citet{mamon00} argues that this $k : \sigma_{\rm cl}$
relation will hold for any well-behaved merger cross section. Given
$\sigma_{\rm cl} \propto T^{1/2}$, for self-similar clusters the merging
rate will thus scale as $k \propto T^{-3/2}$. If we now assume that
the spiral fraction in any particular environment is a function of the
3D density and the merging rate (as is expected given that the number
of mergers endured by a given galaxy per unit time is $P= nk = n
\langle v \Sigma(v) \rangle$ , where $n$ is the 3D number density and 
$\Sigma$ is the velocity dependent merger cross-section, see
\citet{mamon92,mamon00}), we can once again scale the group densities
using the mean cluster and group temperatures, in order to remove the
effects of projection and merging rate. The effects of scaling the
group projected densities by this factor are also shown in
Figure~\ref{fig:spiral_rho}.  This time the group and cluster points
agree fairly well, with only the densest group point departing
somewhat from the cluster trend. Given the likely hierarchical
development of clusters, it is important not to overinterpret this
result. However, it does clearly indicate that at a given 3D density
morphological transformation of galaxies is more effective within
X-ray bright groups rather than within clusters.

We have also separated our group morphology-density relation into
abundances for elliptical and lenticular galaxies. These relations are
compared with the equivalent cluster relations in
Figure~\ref{fig:e_s0_rho}. Note that the group points are plotted at
both the calculated surface densities and the 3D densities adjusted
for their higher expected merging rate (see above). This figure
clearly shows that at equivalent surface densities, the fraction of
both ellipticals and lenticulars in groups tends to be higher than
that in clusters. After scaling the projected densities to allow for
system size and merger rate, as discussed above for spiral fractions,
the points for X-ray bright groups again come into reasonable
agreement with those for clusters, apart from an apparent deficit in
lenticulars in groups (compared to clusters) at the highest densities.
 
\section{Discussion}
\label{sec:dis}

Studies of the evolution of the morphology-density relation (e.g.
\citealt{dressler97}) have shown that the fraction of ellipticals in
clusters has not evolved significantly since z$\sim$0.5. The
elliptical fractions observed in these groups support the suggestion
of \citeauthor{dressler97} that the formation of ellipticals predates
cluster formation, and that they are instead formed primarily in a
group environment. However, the S0 fraction in clusters at z$\sim$0.5
is about 2-3 times smaller than in present day clusters
(e.g. \citealt{dressler97}; although see \citealt{andreon98} for a
different view), suggesting that some process must have transformed
the excess spirals seen in these systems into S0s by the present day.

Major mergers do not appear able to produce enough S0s in clusters
(e.g. \citealt{okamoto01}), and a number of other mechanisms have be
proposed to explain the transformation of spirals to S0s: ram pressure
stripping (e.g. \citealt{gunn72,quilis00}), galaxy harassment
(e.g.\citealt{moore99}) or unequal mass galaxy mergers
(e.g. \citealt{bekki98}).  The fact that we see a comparable S0
fraction in present day X-ray bright groups and clusters, suggests
that whatever process is responsible for converting spirals to S0s in
clusters also plays a significant role in creating S0s in these
groups.

This would appear to rule out ram pressure stripping of the whole ISM
of a galaxy as the dominant process. Indeed, since ram pressure scales
as the square of the galaxy velocity within its group or cluster, ram
pressure stripping should not be effective in the low velocity
dispersion group environment
(e.g. \citealt{zabludoff98,abadi99}). Furthermore, there is also
evidence that ram pressure stripping cannot produce sufficient numbers
of intermediate bulge-to-disk ratio galaxies in clusters
(e.g. \citealt{okamoto01b,lanzoni00}), where it should be more
effective. However, if continuing star formation within spirals is
fuelled by continuing infall of gas from a surrounding gas reservoir
(still a controversial issue), then such a reservoir should be easily
removed by infall into groups as well as clusters, leading to a
gradual decline or `strangulation'
(\citealt{larson80,balogh00,balogh00b}) in star formation.  There is
observational evidence to suggest that this may be occurring, since
late-type galaxies in groups appear to be HI-deficient
(e.g. \citealt{williams87,oosterloo97,verdesmontenegro01}).  The
cessation of star formation would not in itself convert spirals to
lenticulars, so dynamical interactions are probably involved in any
case in effecting structural changes to spirals within dense
environments (tidal disruption is needed to suppress spiral features
and enhance the relative importance of the bulge).

It is tempting to relate the difference in behaviour in the highest
density bin, between the incidence of lenticulars in X-ray bright
groups and clusters, to the apparent evolution in cluster S0 content
since moderate redshifts. In contrast, the elliptical content of
clusters is not found to evolve, and appears to be similar to that of
groups after correction for the higher merger rate in groups. It may
be, for example, that galaxy harassment within the densest regions of
the cluster environment has further boosted the S0 content inherited
from precursor groups: galaxy harassment depends on the frequency of
encounters \citep{moore98b}, which will be higher than the merger
rate, and should be greatest in the centres of clusters.

%%%%%%%%%%%%%%%%%%%%%%%%%%%%%%%%%%%%%%%%%%%%%%%%%%%%%%%%%%%%%%%%%%%%

\section{Acknowledgements}
The authors would like to thank the referee, Gary Mamon, for his many
useful comments and suggestions. This work made use of the Starlink
facilities at Birmingham, the NASA/IPAC Extragalactic Database (NED
--- http://nedwww.ipac.caltech.edu) and the LEDA database
(http://leda.univ-lyon1.fr). SFH acknowledges financial support from
the University of Birmingham.

%%%%%%%%%%%%%%%%%%%%%%%%%%%%%%%%%%%%%%%%%%%%%%%%%%%%%%%%%%%%%%%%%%%%

\bibliography{../bibtex/reffile}

\label{lastpage}

\end{document}